\documentstyle[12pt,epsf]{article}
\newcommand{\beq}{\begin{equation}}
\newcommand{\eeq}{\end{equation}}
\newcommand{\ppd}{\partial}
\newcommand{\epps}{\epsilon}

\title{Chiral Born-Infeld Theory: topological spherically 
symmetrical solitons}

\author{ O.V. Pavlovski\u{\i} \thanks{e-mail address:
ovp@goa.bog.msu.su }  \\ {\em Bogoliubov Institute for Theoretical 
Problems of Microphysics,} \\ {\em Lomonosov Moscow State University 
} \\ {\em Moscow, 119992, Russian Federation.} }

\date{ \ \ \  }

\begin{document}

\maketitle

\begin{abstract}
Finite-energy topological spherically 
symmetrical solutions of  Chiral Born-Infeld theory are studied.  
Properties of these solution are obtained, and  a possible 
physical interpretation is also given.
\end{abstract}

\vspace{1cm}

PACS number(s): 12.39.Dc, 12.39.Fe, 05.45.Yv, 11.27.+d, 11.30.Qc

\section{Introduction}

The concept of the baryon as a chiral  soliton  has a 
long history. The idea of  a unified theory for baryons and mesons that 
is formulated in terms of chiral field  was only proposed for 
the first time by Skyrme
\cite{sk}. The well-known  Skyrme Model is used for analyzing  static
properties of barions, for describing  nucleon-nucleon forces, 
meson-nucleon scattering and for many other low-energy aspects of 
baryons and mesons physics.

In the seventies  't Hooft and Witten \cite{wit}  have shown that,
in the limit $N_c \to \infty$, QCD can be
reformulated   in  terms of effective meson 
theory like Skyrme Model  and that 
chiral solitons of such theory 
reproduce the static properties of real baryons. These facts  
attracted special attention to the theories with chiral solitons.

Effective meson field theory can be  formulated in terms of chiral field
$ U=\exp (i \frac{\vec \phi_\pi \vec \tau}{f_\pi}) $ 
where the vector field $\vec \phi_\pi $ is associated with 
$\pi$ -  mesons and 
$f_\pi = 93 $ MeV is the pion decay constant. The simplest chiral 
invariant prototype lagrangian for mesons field $\vec \phi_\pi $ 
is given by
\beq
{\cal L}_{pr} = - \frac{f^2_\pi}{4} {\rm Tr} L_\mu L^\mu ,
\label{1.1}
\eeq
where $L_\mu=U^+ \ppd_\mu U$  is a Cartan left-invariant form (see reviews 
\cite{rev}).  However using the scale transformation analogous to the Derrick 
Theorem \cite{Derik} one can find that theory with the lagrangian (\ref{1.1}) has 
no finite energy static solutions and all models with chiral soliton 
appear to be  
various modifications of the prototype lagrangian. For example, the 
skyrmeon (chiral soliton ofthe  Skyrme Model) exists due to the including of 
the
Skyrme-part $\Delta {\cal L}_{Sk}={\rm Tr} [L_\mu,L_\nu ]^2/e^2$ in 
lagrangian of this model. 
This theory has a set of topological static solutions and one can classify 
these solutions by the value of the topological (baryon) charge 
\beq
B = \frac{1}{24\pi^2} \int d^3 \epps_{ijk} {\rm Tr} (L_i L_j L_k ) .
\label{1.2} 
\eeq
In this 
model such solutions are associated with baryon states with different baryon 
charges, and soliton with $B=1$ (skyrmeon) is treated as a nucleon. Such kind 
of solutions and quantum fluctuations about them is well studied now 
(see \cite{rev}) and results of such analysis are in agreement with the 
experiment 
data with accuracy of 30-40 percents for baryonic masses and another static 
properties of baryons.

But there are many other ways for stabilization of chiral soliton, because
from the methodological point of view, arising of the Skyrme part in the
lagrangian of effective meson field theory is an ''ad hoc" procedure, and 
there 
are no real physical bases for this procedure. Of course, one can treat 
this part in the 
effective lagrangian  as a leading high derivative expansion of 
effective chiral meson action (see review \cite{ioffe}), but this approach 
leads to numerous questions so far. Two most important ones concern the 
physical nature of the scale parameter ($e$) in this theory and the 
influence 
of another terms of expansion  on existence and stability of chiral 
solitons.

Another possible way of stabilizing of chiral solitons was suggested in
\cite{omega}. In these works the role of the vector ($\rho$) and 
pseudo-scalar ($\omega$) mesons in chiral soliton physics is discussed.
Procedure of $\omega$-stabilization leads to forming of topological soliton
solution. Many other procedures of stabilization were proposed
so far. One such approach that allows to get over the difficulty
connected with scale instability of chiral solitons in 3+1 dimension was 
proposed in \cite{deser}.
Indeed, prototype lagrangian (\ref{1.1}) has well-known stable static
topological solitons in 1+1 dimension \cite{pol}. This situation stems from the fact 
of dimensionless of the chiral action in 1+1 dimension. 
By analogy with 1+1 dimension case, in \cite{deser} the model with the 
lagrangian
\beq
{\cal L}_{3/2} = - \frac{f^2_\pi}{4} ({\rm Tr} L_\mu L^\mu)^{3/2} ,
\label{1.3}
\eeq
was considered. The authors of \cite{deser2}  name this model  
Born-Infeld-Skyrme Model because of
the half-integer power in expression (\ref{1.3}). 

In \cite{Dion} another type of a Born-Infeld-Skyrme Model with the lagrangian
\beq
{\cal L}_{SkBI} = - \frac{f^2_\pi}{4} {\rm Tr} L_\mu L^\mu 
 +\beta^2\left( \sqrt{1+\frac 1{16e^2\beta^2}
{\rm Tr}
\left[ L_{\mu ,},L_\nu \right] ^2}-1\right)
\label{1.4}
\eeq
was proposed. This model satisfies the basic requirements for models of
Born-Infeld
type and have stable soliton solution, but the Born-Infeld part in the 
lagrangian
(\ref{1.4}) plays the role of Skyrme-term in the lagrangian (\ref{1.3}) 
and this 
model is one possible modification of Skyrme's procedure.

The consistent effective low-energy meson's theory construction is a very 
complicated task that is closely connected to the quark's confinement 
problem and to the problem of spontaneous breaking of chiral 
symmetry. It is desirable that such theory should be 
Lorentz and chiral invariant and should have a set of  finite
energy stable solutions. In the case of electro-magnetic fields such kind 
of an effective low-energy theory is the well-known Born-Infeld model
\cite{born}. In this work we study the direct analogue of the Born-Infeld 
action for  chiral fields. 
By construction our model has no singular solutions and looks 
very attractive as a possible effective action for mesons fields. 
This model has a set of stable topological solitons. These solutions can be
treated as baryons states in our model. The physical motivation of our model 
is beyond the scope of this article.

In Section 2, we discuss the problem of constructing of a Born-Infeld (BI)
action for the chiral field (ChBI model). 
The lagrangian and equation of motion 
are presented. The spherically symmetrical configurations of 
chiral field are considered  and topological spherically symmetrical solitons 
are studied. In Section 3 we summarise the results of this paper.

\section{The Born-Infeld theory for Chiral Fields }

In the  well-known paper \cite{born}, Born and Infeld  proposed a non-linear
covariant action for electro-magnetic fields with very attractive features.
Firstly, in the framework of BI theory the problem of singular self-energy of
electron can be solved. In this theory the electron is a stable finite energy
solution of the BI field equation with electric charge.
Second, the BI action has a scale parameter $\beta$. Using expansion by this
parameter one reduces the BI action to the usual Maxwell form in the low-energy
limit.

We want  to perform a very similar procedure with the chiral prototype
lagrangian (\ref{1.1}).
Like in the case of BI action, our chiral model must have a set of  finite
energy solutions with integer values of charge (topological or baryon), and
in the low-energy limit such a theory must reproduce the prototype 
lagrangian (\ref{1.1}). The model must be 
Lorentz and chiral invariant. Finally, the
form of such theory follows from the analogy with the action of a 
relativistic 
particle, the BI action for EM and YM field. This form has geometrical nature 
and corresponds to a number of analogies between chiral dynamics, gauge 
theory and general relativity \cite{pak}.

Arguing as above, let us consider a theory with lagrangian
\beq
{\cal L}_{ChBI} = - f^2_\pi {\rm Tr} 
\beta^2 \bigg(1-\sqrt{1-\frac 1{2\beta^2}L_\mu L^\mu } \bigg) 
\sim 
- \frac{f^2_\pi}{4} {\rm Tr} L_\mu L^\mu ,
\label{2.1}
\eeq
where $\beta$ is a mass dimensional scale parameter of our model. It is
easily shown that the expansion of the lagrangian (\ref{2.1}) gives us 
the prototype 
theory as the leading order theory by parameter $\beta$. Our theory contains 
only second-order derivative terms. Thus the dynamics of this field theory 
can be studied in detail.

Now we consider the spherically symmetrical field
configuration
\beq
U=e^{i F(r) (\vec n \vec \tau)}, \qquad \vec n = \vec r/|r| .
\label{2.2}
\eeq
The energy of such field configuration is the functional
\beq
E^\beta [F] = 
8 \pi f^2_\pi \beta^2 \int\limits_0^\infty
\Big( 1 - R \Big) r^2 dr , 
\label{2.5} 
\eeq
where
$$
R=\sqrt{ 
1-\frac 1{\beta^2}\Big(\frac{F'^2}{2} +  \frac {\sin^2 F}{r^2}  \Big)
 } .
$$

Using the variation principle, we get the
equation of motion
\beq
\Big(r^2 \frac{F'}{R} \Big)' =  \frac{\sin 2F}{R}
\label{2.3}
\eeq
and for amplitude $F(r)$ we obtain
$$
(r^2 - \frac 1{\beta^2} \sin^2 F) F'' + ( 2rF' -\sin 2F) -
$$
\beq
- \frac 1{\beta^2} (r F'^3 - F'^2 \sin 2F + 3 \frac 1{r} F' \sin^2F 
- \frac 1{r^2} \sin2F \sin^2 F)=0.
\label{2.4}
\eeq

 \begin{figure}[t]
 \leavevmode
 \epsfbox{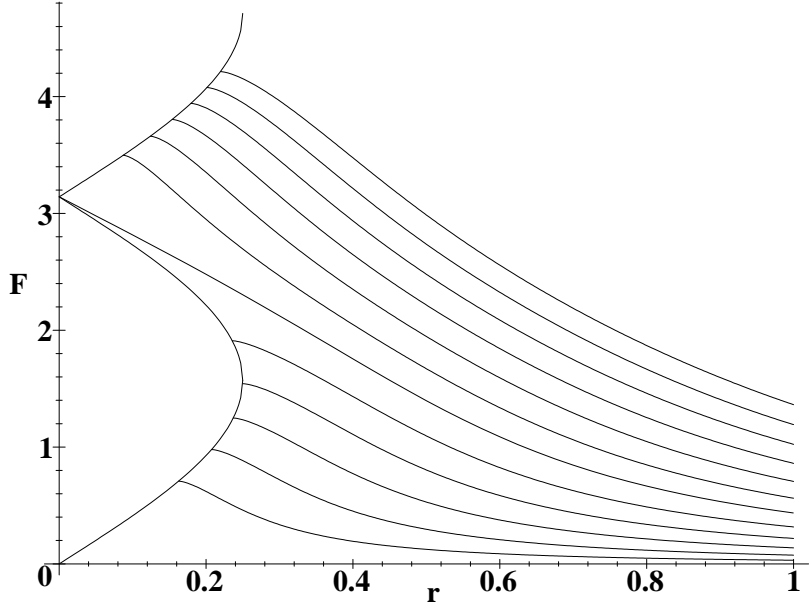}
 \caption{ Solutions of equation (9) 
which have the asymptotics (12) ($b>0$) at infinity. 
Horizontal axis: r (in fm).}
 \end{figure}

The next aim of our investigation is finding the solutions of
equation (\ref{2.4}). 

Equation (\ref{2.4}) is a very complicated nonlinear differential
equation. In order to solve it, only numerical or approximation
methods seem applicable. The crucial point of such analysis is that
the leading derivative term in this equation contains the factor
\beq
\Phi [F] (r) = \Bigl( r^2 - \frac 1{\beta^2} \sin^2 F \Bigr).
\label{2.6}
\eeq
If there exits $r_0 \neq 0$ such that $F(r_0)= \pm \arcsin (\beta r_0)$, 
then it is easily shown that $F'(r)|_{r=r_0}=0$. 
But if $r_0 = 0$, then $F(0)=\pi N$ where $N \in {\bf Z}$ and 
$F'(r)|_{r=0}=a \neq 0 $. 
Using the standard procedure, one obtains the 
asymptotic behavior near this point ($r=0$, $F(0)=\pi N$) 
\beq
F(r) = \pi N  + a r - \frac{7a^2-4\beta^2}{30(a^2-\beta^2)} a^3  
r^3+ \underline{O}(r^{5}),  
\label{2.7} 
\eeq 
where $a^2 < \beta^2/3$ is a constant.  Of course, the 
derivative at the point $r=R$ is not zero.  

To guarantee the stability of our solutions we should choose 
the following asymptotics at infinity ($r \to \infty$)
\beq
F(r) =  b \, (1/r)^2  - \frac{b^3}{21} \, (1/r)^6 -
\frac{b^3}{3\beta^2}  \, (1/r)^8 +
\underline{O}(1/r^{10}), 
\label{2.8}
\eeq
where $b$ is a constant.
Solutions with such asymptotics at infinity and
with asymptotics (\ref{2.7}) ($N \neq 0$) are stable because vacuum states
$F(0)= \pi N$ and $F(\infty)=0$ are different.

Notice that equation (\ref{2.4}) has very useful symmetries.
First of all, this equation is symmetrical with respect to the
changes $F \leftrightarrow F+N\pi, \, N \in {\bf Z} $ and 
$F \leftrightarrow -F$.

 \begin{figure}[t]
 \leavevmode
 \epsfbox{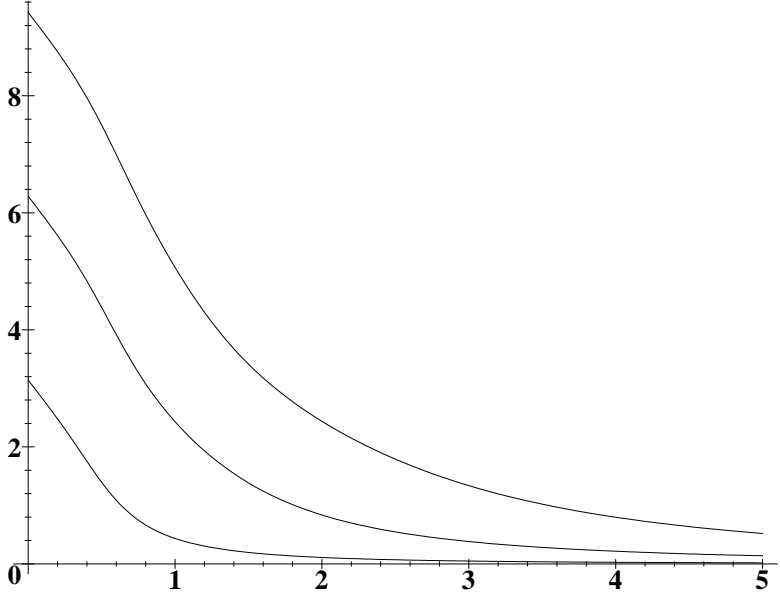}
 \caption{Solitons with $B=1, 2$ and $3$.
  Horizontal axis: r (in fm).}
 \end{figure}

\begin{table}[thb]
\begin{center}
\begin{tabular}{|c|c||c|c|}
\hline
Baryon Charge,  & Energy of Soliton, & 
Baryon Charge,  & Energy of Soliton, \\
  $B$  &   $E(\beta=1)$ & $B$  &   $E(\beta=1)$ \\
\hline
1 & 3.487  & 6  &  300.08   \\
\hline
2 & 17.069  & 7  & 459.88    \\
\hline
3 & 47.150  & 8  & 668.05    \\
\hline
4 & 100.10  & 9  & 930.97    \\
\hline
5 & 182.28  & 10 & 1255.01    \\
\hline
\end{tabular} 
\end{center}
\caption{Dependence of the energy $E( B, \, \beta=1)$ from 
value of Baryon Charge $B$}
\end{table}


The numerical investigation of the solutions of equation (\ref{2.4}) 
which have the asymptotics (\ref{2.8}) ($b>0$) at infinity is
presented in Fig.1. Most of such solutions can be  evaluated only
for $r>r_0$, where $r_0$ is determined by $F(r_0)= \pm \arcsin (\beta r_0)$. 
But among this set of solutions there are solutions with $r_0=0$. 
Such solutions have the
asymptotics (\ref{2.7}) at origin, and these are the topological solitons of
the ChBI theory. The topological charge is defined by integral (\ref{1.2}) as usual.
Solitons with $B=1, 2$ and $3$ are presented in Fig.2. The scale parameter
$\beta=807 \, \mbox{MeV}$ is defined from the hypothesis that the soliton 
with 
$B=1$ is a nucleon. Indeed, using now a scale transformation in (\ref{2.5}), 
one gets 
\beq
\beta = \frac{8 \pi f^2_\pi }{m_p} \, E( B=1, \, \beta=1)=807 \, 
\mbox{MeV} ,
\label{2.10} 
\eeq
where $E( B=1, \, \beta=1)=3.487$ is the energy of this soliton solution 
for $B=1$ and $\beta=1$. 

In conclusion of this section let us consider the question about the 
asymptotics 
of energy of spherically symmetrical solitons at $B \to \infty$. 
In case of the Skyrme Model $E(B) \sim B^2$ \cite{Bog}. 
In the Chiral-Born-Infeld's case it is not so. In Table 1 the values of the 
energy $E( B, \, \beta=1)$ for $B=1 \ldots 10$ are presented.

 \begin{figure}[thb]
 \leavevmode
 \epsfbox{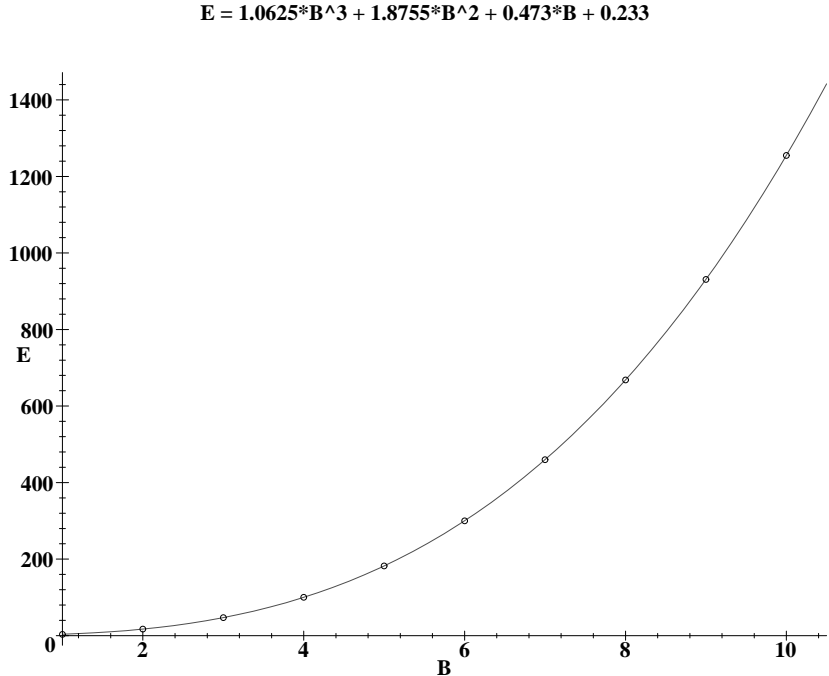}
 \caption{The energy $E( B, \, \beta=1)$: 
           interpolation of data of Table 1. }
 \end{figure}

It is easily shown that the Table 1 data can be interpolated by 
the expression
\beq
E( B, \, \beta=1) \simeq 1.0625 B^3 + 1.8755 B^2 + 0.473 B + 0.233 .
\label{2.11} 
\eeq  
This interpolation is shown in Fig.3 where dots are the data from Table 1. 

\section{Conclusions}
The aim of this paper is to construct a Born-Infeld theory for 
chiral fields. This theory has no singular solution and so the Chiral 
Born-Infeld theory is a good candidate on the role of effective chiral theory. 
In our work we show that this theory has the stable
cluster finite-energy solution. In the concept of a unified theory of mesons
and baryons \cite{sk} such solutions are treated as the baryon states.

Of course, we do not give a comprehensive investigation of  the
Born-Infeld theory. The questions about non-spherical solutions,
quantum fluctuations about such solutions and corresponding 
properties of baryons or about the nucleon-nucleon interaction  
are not clear now. But maybe the most
important question in such investigation is 
about physical substantiation of such theory. All of these questions 
should be the themes for a future investigation.

In conclusion I would like to draw attantion to another class of 
solutions that 
were studied in Section 2. These solutions are defined everywhere, except
the little ($ \sim$ 0.2 fm) spherical region about the origin. These 
solutions 
look like a ''bubble" of vacuum in the chiral fields and are of interest for 
the chiral bag model of baryons \cite{bag}. This question 
should be the theme for a future investigation too.

\end{document}